%% file: ce_over_arxiv.tex
\documentclass{article}
\usepackage{spconf,amsmath,graphicx}

\usepackage{tikz}
\usetikzlibrary{dsp,chains,arrows}
\usepackage{caption}
\usepackage{enumitem}
\tikzset{
    font={\fontsize{9}{11.0476pt}\selectfont}}
\usepackage{amsmath}
\usepackage{mathtools,amssymb}
\usepackage{pgfplots}
\pgfplotsset{compat=newest}

\title{Investigation of Channel Estimation Techniques with 1-bit Quantization and Oversampling for Multiple-Antenna Systems}

\name{Zhichao~Shao, Lukas~T.~N.~Landau and Rodrigo~C.~de~Lamare}
\address{Centre for Telecommunications Studies\\
    Pontifical Catholic University of Rio de Janeiro,
    Rio de Janeiro, Brazil 22453-900\\
    Email: zhichao.shao;lukas.landau;delamare@cetuc.puc-rio.br}

\begin{document}
\ninept
\maketitle

\begin{abstract}
Large-scale multiple-antenna systems have been identified as a promising technology for the next generation of wireless systems. However, by scaling up the number of receive antennas the energy consumption will also increase. One possible solution is to use low-resolution analog-to-digital converters at the receiver. This paper considers large-scale multiple-antenna uplink systems with 1-bit analog-to-digital converters on each receive antenna. Since oversampling can partially compensate for the information loss caused by the coarse quantization, the received signals are firstly oversampled by a factor M. We then propose a low-resolution aware linear minimum mean-squared error channel estimator for 1-bit oversampled systems. Moreover, we characterize analytically the performance of the proposed channel estimator by deriving an upper bound on the Bayesian Cram\'er-Rao bound. Numerical results are provided to illustrate the performance of the proposed channel estimator.
\end{abstract}

\begin{keywords}
Large-scale multiple-antenna systems, 1-bit quantization, oversampling, channel estimation, Bayesian CRB
\end{keywords}

\section{Introduction}

With a large number of receive antennas at the base station (BS)
massive multiple-input-multiple-output (MIMO) systems can
significantly increase the spectral efficiency,  mitigate the
propagation loss caused by channel fading, reduce the
inter-user-interference and have many other advantages as compared
to current systems \cite{Larsson,6375940}. Despite all these
benefits, massive MIMO has brought some challenges. For example, by
using current high-resolution analog-to-digital converters (ADCs)
for each element of the antenna arrays at the BS the hardware cost
and the energy consumption may become prohibitively high. To address
this challenge, low-cost and low-resolution ADCs are promoted.

Many works have studied massive MIMO systems \cite{mmimo,wence} with
low-resolution ADCs (e.g. 1-3 bits) at the front-end. The authors in
\cite{7894211,7458830,6987288} have investigated the uplink
performance by multiple-user (MU) massive MIMO systems using ADCs
with only a few bits of resolution. Millimeter-Wave (mmWave) massive
MIMO systems are favorable candidates for the next generation
cellular systems. The major benefit is that they can achieve much
larger bandwidths. The authors in \cite{8337813,7094595,6804238}
have discussed channel estimation, signal detection, achievable rate
and precoding techniques for mmWave massive MIMO systems with
low-resolution ADCs at the radio frequency (RF) chains. As one
extreme case, 1-bit ADCs can dramatically decrease the energy
consumption of the receiver. Recent studies include precoding
\cite{8010806}, channel estimation \cite{8385500}, capacity analysis
\cite{7155570} and iterative detection and decoding (IDD) techniques
\cite{8240730}. In order to mitigate the performance loss caused by
coarse quantization, oversampling is a common used technique, where
the received signal is sampled at a rate faster than the Nyquist
rate \cite{7837644,LandauTWC}. The work in \cite{7996526} has
proposed an oversampling technique to obtain better multiuser
interference suppression and error rate performance. To further
reduce the computational complexity caused by the inversion of a
large matrix in oversampled system, a sliding window based linear
detection is proposed in \cite{8450809}.

Currently, channel estimation is a known problem that limits the performance of 1-bit ADCs systems. In this paper, we investigate channel estimation techniques for uplink 1-bit and oversampled MIMO systems. One essential and unique aspect of our proposed channel estimator is that oversampling is taken into account, which can significantly improve the performance. In particular, we develop a low-resolution aware (LRA) linear minimum mean-squared error (LMMSE) channel estimator for 1-bit oversampled systems based on the Bussgang decomposition. Unlike the proposed channel estimator in \cite{8487039}, we consider the correlation of the filtered noise, which is important for the oversampled system. We also examine the fundamental estimation limits by deriving a Bayesian framework for both non-oversampled and oversampled systems.

The rest of this paper is organized as follows: Section 2 shows the system model and gives some statistical properties of 1-bit quantization. Section 3 illustrates the Bayesian information for 1-bit non-oversampled and oversampled MIMO systems and gives a short derivation of the proposed oversampling based LRA-LMMSE channel estimator. In section 4, the simulation results are presented and section 5 concludes the paper.

The following notations are used: matrices are in bold capital letters while vectors in bold lowercase. $\mathbf{I}_n$ denotes $n\times n$ identity matrix. $\mathbf{0}_n$ is a $n\times 1$ all zeros column vector. Additionally, $\text{diag}(\mathbf{A})$ is a diagonal matrix only containing the diagonal elements of $\mathbf{A}$. The vector or matrix transpose and conjugate transpose are represented by $(\cdot)^T$ and $(\cdot)^H$, respectively.  $[\cdot]_k$ represents the $k$th element of the corresponding vector. $(\cdot)^R$ and $(\cdot)^I$ gets the real and imaginary part from the corresponding vector or matrix, respectively. $\otimes$ is the Kronecker product and $\det(\cdot)$ is the determinant function.

\begin{figure*}[!htbp]
    \centering
    \input{system_model.tex}
    \caption{System model of multi-user multiple-antenna system with 1-bit ADCs and oversampling at the receiver}
    \label{fig:transmitter}
    \vspace{-0.4cm}
\end{figure*}
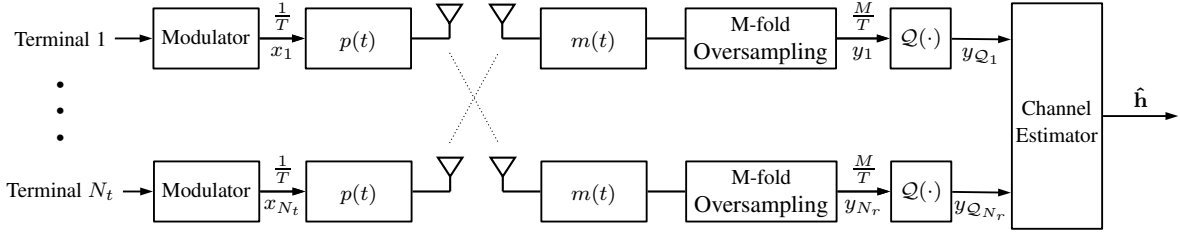

\section{System Model}
We consider the uplink of a single-cell multi-user large-scale multiple-antenna system, which is shown in Fig. 1. At the transmitter side there are $N_t$ single-antenna terminals, whereas $N_r$ receive antennas are employed at the BS. For the large-scale MIMO system we have $N_r\gg N_t$. With perfect synchronization the received oversampled signal $\mathbf{y}\in\mathbb{C}^{MN_rN \times 1}$ can be expressed as
\begin{equation}
\mathbf{y}=\mathbf{H}\mathbf{x}+\mathbf{n},
\label{equ:system_model}
\end{equation}
where $\mathbf{x}\in\mathbb{C}^{NN_t \times 1}$ contains independent identically distributed (i.i.d.) transmitted symbols from $N_t$ terminals, each with block length $N$. Each symbol has unit energy so that $E\{\mathbf{x}_k^2\}=1$. The vector $\mathbf{n}$ represents the filtered oversampled noise expressed by
\begin{equation}
\mathbf{n} = \left(\mathbf{I}_{N_r}\otimes \mathbf{G}\right)\mathbf{w}
\end{equation}
with $\mathbf{w}\sim \mathcal{CN}\left(\mathbf{0}_{3MN_rN},\sigma^2_n\mathbf{I}_{3MN_rN}\right)$. Note that the noise samples are described such that each entry of $\mathbf{n}$ has the same statistical properties. Since the receive filter has a length of $2MN + 1$ samples, $3MN$ unfiltered noise samples in the noise vector $\mathbf{w}$ need to be considered for the description of an interval of $MN$ samples of the filtered noise $\mathbf{n}$. $\mathbf{G}\in\mathbb{R}^{MN \times 3MN}$ is a Toeplitz matrix that contains the coefficients of the matched filter $m(t)$ at different time instants:
    \begin{equation}
    \resizebox{.48\textwidth}{!}{$\displaystyle
    \mathbf{G} = \begin{bmatrix}
    m(-NT)& m(-NT+\frac{1}{M}T)& \dots& m(NT) & 0 & \dots & 0\\
    0 & m(-NT)& \dots & m(NT-\frac{1}{M}T) & m(NT) & \dots & 0\\
    \vdots & \vdots & \ddots & \vdots & \vdots & \ddots & \vdots\\
    0 & 0 & \dots & m(-NT)& m(-NT+\frac{1}{M}T)& \dots& m(NT)\\
    \end{bmatrix}.$}
    \label{equ:R}
    \end{equation}
$T$ is the symbol period and $M$ denotes the oversampling rate. The equivalent channel matrix $\mathbf{H}$ is described as
\begin{equation}
\mathbf{H} = \left(\mathbf{I}_{N_r}\otimes \mathbf{Z}\right)\mathbf{U}\left(\mathbf{H}'\otimes \mathbf{I}_N\right),
\label{equ:H}
\end{equation}
where $\mathbf{H}'\in\mathbb{C}^{N_r \times N_t}$ is the channel matrix for non-oversampled system and $\mathbf{U}$ is an oversampling matrix, which can be calculated as
\begin{equation} 
\mathbf{U} = \mathbf{I}_{N_rN}\otimes\mathbf{u} = \mathbf{I}_{N_rN}\otimes \left[0 \quad \cdots \quad 0 \quad 1\right]^T_{1 \times M}.
\end{equation}
$\mathbf{Z}\in\mathbb{R}^{MN \times MN}$ is a Toeplitz matrix that contains the coefficients of $z(t)$ at different time instants, where $z(t)$ is the convolution of the pulse shaping filter $p(t)$ and the matched filter $m(t)$, and is given by
\begin{equation}
\resizebox{.47\textwidth}{!}{$\displaystyle
    \mathbf{Z} = \begin{bmatrix}
    z(0) & z(\frac{T}{M}) & \dots & z(NT-\frac{1}{M}T)\\
    z(-\frac{T}{M}) & z(0) & \dots & z(NT-\frac{2}{M}T)\\
    \vdots & \vdots & \ddots & \vdots\\
    z(-NT+\frac{1}{M}T) & z(-NT+\frac{2}{M}T) & \dots & z(0)\\
    \end{bmatrix}.$}
\end{equation}
In particular, $M=1$ refers to the non-oversampling case.

Let $\mathcal{Q}(\cdot)$ represent the 1-bit quantization function, the resulting quantized signal $\mathbf{y}_\mathcal{Q}$ is
\begin{equation}
\mathbf{y}_\mathcal{Q}=\mathcal{Q}\left(\mathbf{y}\right)=\mathcal{Q}(\mathbf{y}^R) + j\mathcal{Q}(\mathbf{y}^I).
\label{system_model}
\end{equation}
The real and imaginary part of $\mathbf{y}$ are element-wisely quantized to $\{\pm\frac{1}{\sqrt{2}}\}$ based on the sign.

Since quantization strongly changes the properties of signals, some statistical properties of quantization for a Gaussian input signal will be shown. For 1-bit quantization and Gaussian inputs, the cross-correlation between the unquantized signal $\mathbf{s}$ with covariance matrix $\mathbf{C}_\mathbf{s}$ and its 1-bit quantized signal $\mathbf{s}_\mathcal{Q}$ is described by \cite{Bussgang}
\begin{equation}
\mathbf{C}_{\mathbf{s}_\mathcal{Q}\mathbf{s}}=\sqrt{\frac{2}{\pi}}\mathbf{K}\mathbf{C}_{\mathbf{s}},\mbox{where } \mathbf{K}=\text{diag}(\mathbf{C}_{\mathbf{s}})^{-\frac{1}{2}}.
\end{equation}
Furthermore, the covariance matrix of the 1-bit quantized signal $\mathbf{s}_\mathcal{Q}$ can be obtained through the arcsin law \cite{Jacovitti} as follows:
\begin{equation}
\mathbf{C}_{\mathbf{s}_\mathcal{Q}}=\frac{2}{\pi}\left(\text{sin}^{-1}\left(\mathbf{K}\mathbf{C}_{\mathbf{s}}^R\mathbf{K}\right)+j\text{sin}^{-1}\left(\mathbf{K}\mathbf{C}_{\mathbf{s}}^I\mathbf{K}\right)\right).
\label{equ_arcsin}
\end{equation}

\section{Channel Estimation for 1-Bit MIMO}
The problem of interest here is to estimate the channel parameters in $\mathbf{H}'$ from the received quantized signal $\mathbf{y}_\mathcal{Q}$. In the following, we firstly derive the Bayesian Cram\'er-Rao bound (CRB) in terms of the Bayesian information for both non-oversampled and oversampled systems. Thereafter, we make a short derivation of the proposed oversampling based LRA-LMMSE channel estimator.

Through vectorization, the received signal in (\ref{equ:system_model}) is
\begin{equation}
\begin{aligned}
\mathbf{y}&=(\mathbf{x}^T\otimes\mathbf{I}_{N_rNM})\text{vec}(\mathbf{H})+\mathbf{n}\\&=[\mathbf{x}^T\otimes\mathbf{I}_{N_r}\otimes\mathbf{Z}(\mathbf{I}_{N}\otimes\mathbf{u})]\text{vec}(\mathbf{H}'\otimes \mathbf{I}_N)+\mathbf{n}.
\end{aligned}
\label{equ_sys}
\end{equation}
To further simplify $\text{vec}(\mathbf{H}'\otimes \mathbf{I}_N)$, we have
\begin{equation}
\begin{aligned}
\text{vec}(\mathbf{H}'\otimes \mathbf{I}_N) &= \\& \hspace{-1.4cm} \left[\mathbf{I}_{N_t}\otimes\left(\mathbf{e}_1\otimes\mathbf{I}_{N_r}\otimes\mathbf{e}_1+\dots+\mathbf{e}_N\otimes\mathbf{I}_{N_r}\otimes\mathbf{e}_N\right)\right]\text{vec}(\mathbf{H'}),
\end{aligned}
\end{equation}
where $\mathbf{e}_n$ represents an all zeros column vector except for the $n$th element which is one. Then (\ref{equ_sys}) can be summarized as
\begin{equation}
\mathbf{y}=\mathbf{\Phi}\text{vec}(\mathbf{H'})+\mathbf{n}=\mathbf{\Phi}\mathbf{h'}+\mathbf{n},
\label{Equ_nonquan}
\end{equation}
where $\mathbf{\Phi}$ is the equivalent transmit matrix. The channel parameters in $\mathbf{h'}$ are assumed to be random complex Gaussian distributed with zero mean and covariance matrix $\mathbf{C}_{\mathbf{h'}}$.

\subsection{Bayesian Bounds on Channel Estimation}
We rewrite the complex-valued system (\ref{Equ_nonquan}) in the following real-valued form
\begin{equation}
\begin{bmatrix}
\mathbf{y}^R \\
\mathbf{y}^I
\end{bmatrix}=\begin{bmatrix}
\mathbf{\Phi}^R &-\mathbf{\Phi}^I\\
\mathbf{\Phi}^I & \mathbf{\Phi}^R
\end{bmatrix}\begin{bmatrix}
\mathbf{h'}^R \\
\mathbf{h'}^I
\end{bmatrix}+\begin{bmatrix}
\mathbf{n}^R \\
\mathbf{n}^I
\end{bmatrix}.
\label{equ:system_model_real}
\end{equation}
Considering the unknown parameter vector $\tilde{\mathbf{h'}}=[\mathbf{h'}^R;\mathbf{h'}^I]$, since the real and imaginary parts are independent, the Bayesian information matrix (BIM) \cite{van2013detection} is defined as
\begin{equation}
\mathbf{J}_{\mathbf{y}_\mathcal{Q}}(\tilde{\mathbf{h'}})=\mathbf{J}_{\mathbf{y}_\mathcal{Q}^R}(\tilde{\mathbf{h'}})+\mathbf{J}_{\mathbf{y}_\mathcal{Q}^I}(\tilde{\mathbf{h'}}),
\label{equ_total_baysian}
\end{equation}
where
\begin{equation}
    [\mathbf{J}_{\mathbf{y}_\mathcal{Q}^{R/I}}(\tilde{\mathbf{h'}})]_{ij}=E_{\mathbf{y}_\mathcal{Q}^{R/I},\tilde{\mathbf{h'}}}\left\{\frac{\partial \ln p(\mathbf{y}_\mathcal{Q}^{R/I},\tilde{\mathbf{h'}})}{\partial [\tilde{\mathbf{h'}}]_i}\frac{\partial \ln p(\mathbf{y}_\mathcal{Q}^{R/I},\tilde{\mathbf{h'}})}{\partial [\tilde{\mathbf{h'}}]_j}\right\},
    \label{equ_bayesian}
\end{equation}
with $[\tilde{\mathbf{h'}}]_i$ and $[\tilde{\mathbf{h'}}]_j$ being the elements of $\tilde{\mathbf{h'}}$ and $\mathbf{J}_{\mathbf{y}_\mathcal{Q}}(\tilde{\mathbf{h'}})$ is arranged as follows:
\begin{equation}
\mathbf{J}_{\mathbf{y}_\mathcal{Q}}(\tilde{\mathbf{h'}})=\begin{bmatrix}
[\mathbf{J}_{\mathbf{y}_\mathcal{Q}}(\tilde{\mathbf{h'}})]_{RR} & [\mathbf{J}_{\mathbf{y}_\mathcal{Q}}(\tilde{\mathbf{h'}})]_{RI} \\ [\mathbf{J}_{\mathbf{y}_\mathcal{Q}}(\tilde{\mathbf{h'}})]_{IR} & [\mathbf{J}_{\mathbf{y}_\mathcal{Q}}(\tilde{\mathbf{h'}})]_{II}
\end{bmatrix}.
\end{equation}
Eq.(\ref{equ_bayesian}) can be divided into two parts
\begin{equation}
    [\mathbf{J}_{\mathbf{y}_\mathcal{Q}^{R/I}}(\tilde{\mathbf{h'}})]_{ij} = [\mathbf{J}^D_{\mathbf{y}_\mathcal{Q}^{R/I}}(\tilde{\mathbf{h'}})]_{ij}+[\mathbf{J}^P_{\mathbf{y}_\mathcal{Q}^{R/I}}(\tilde{\mathbf{h'}})]_{ij},
\end{equation}
where
\begin{equation}
    [\mathbf{J}^D_{\mathbf{y}_\mathcal{Q}^{R/I}}(\tilde{\mathbf{h'}})]_{ij}\triangleq E_{\mathbf{y}_\mathcal{Q}^{R/I}\mid\tilde{\mathbf{h'}}}\left\{\frac{\partial \ln p(\mathbf{y}_\mathcal{Q}^{R/I}\mid\tilde{\mathbf{h'}})}{\partial [\tilde{\mathbf{h'}}]_i}\frac{\partial \ln p(\mathbf{y}_\mathcal{Q}^{R/I}\mid\tilde{\mathbf{h'}})}{\partial [\tilde{\mathbf{h'}}]_j}\right\}
    \label{equ_Jd}
\end{equation}
and
\begin{equation}
    [\mathbf{J}^P_{\mathbf{y}_\mathcal{Q}^{R/I}}(\tilde{\mathbf{h'}})]_{ij} \triangleq E_{\tilde{\mathbf{h'}}}\left\{\frac{\partial \ln p(\tilde{\mathbf{h'}})}{\partial [\tilde{\mathbf{h'}}]_i}\frac{\partial \ln p(\tilde{\mathbf{h'}})}{\partial [\tilde{\mathbf{h'}}]_j}\right\}.
    \label{equ_h}
\end{equation}
To transform the real value $\tilde{\mathbf{h'}}$ back to the complex domain $\mathbf{h'}$, we apply the chain rule to get:
\begin{equation}
\begin{aligned}
\mathbf{J}_{\mathbf{y}_\mathcal{Q}}(\mathbf{h'})&=\frac{1}{4}\left([\mathbf{J}_{\mathbf{y}_\mathcal{Q}}(\tilde{\mathbf{h'}})]_{RR}+[\mathbf{J}_{\mathbf{y}_\mathcal{Q}}(\tilde{\mathbf{h'}})]_{II}\right)\\&+\frac{j}{4}\left([\mathbf{J}_{\mathbf{y}_\mathcal{Q}}(\tilde{\mathbf{h'}})]_{RI}-[\mathbf{J}_{\mathbf{y}_\mathcal{Q}}(\tilde{\mathbf{h'}})]_{IR}\right).
\end{aligned}
\end{equation}
The variance of the LMMSE estimator $\hat{\mathbf{h'}}(\mathbf{y}_\mathcal{Q})$ is lower bounded by
\begin{equation}
    \text{var}[\hat{h'}_i(\mathbf{y}_\mathcal{Q})]\geq[\mathbf{J}_{\mathbf{y}_\mathcal{Q}}^{-1}(\mathbf{h'})]_{ii}.
\end{equation}

\subsubsection{BIM for Non-oversampled Systems}
For non-oversampled systems, i.e. $M=1$, the covariance matrix of the equivalent noise vector $\mathbf{n}$ is $\mathbf{C}_\mathbf{n}=\sigma_n^2\mathbf{I}_{NN_r}$. Since $\mathbf{n}$ is white noise, the conditional log-likelihood function can be expressed as
\begin{equation}
\begin{aligned}
\ln p(\mathbf{y}_\mathcal{Q}\mid\tilde{\mathbf{h'}})&=\sum_{k=1}^{NN_r}\left[\ln p([\mathbf{y}_\mathcal{Q}^R]_k\mid[\tilde{\mathbf{h'}}]_k)+\ln p([\mathbf{y}_\mathcal{Q}^I]_k\mid[\tilde{\mathbf{h'}}]_k)\right]\\&\hspace{-0cm}=\sum_{k=1}^{NN_r}\ln Q\left(-\sqrt{2}[\mathbf{y}_\mathcal{Q}^R]_k\frac{[\mathbf{\Phi}^R\mathbf{h'}^R-\mathbf{\Phi}^I\mathbf{h'}^I]_k}{\sigma_n/\sqrt{2}}\right)\\&+\sum_{k=1}^{NN_r}\ln Q\left(-\sqrt{2}[\mathbf{y}_\mathcal{Q}^I]_k\frac{[\mathbf{\Phi}^I\mathbf{h'}^R+\mathbf{\Phi}^R\mathbf{h'}^I]_k}{\sigma_n/\sqrt{2}}\right),
\end{aligned}
\label{equ_llr}
\end{equation}
where $Q(x) = \frac{1}{\sqrt{2\pi}}\int_x^\infty \exp(-\frac{u^2}{2})du$. Inserting (\ref{equ_llr}) into (\ref{equ_Jd}) we obtain
 \begin{equation}
    [\mathbf{J}^D_{\mathbf{y}_\mathcal{Q}}(\tilde{\mathbf{h'}})]_{ij} = -E\left\{\frac{\partial^2\ln p(\mathbf{y}_\mathcal{Q}\mid\tilde{\mathbf{h'}})}{\partial [\tilde{\mathbf{h'}}]_i\partial [\tilde{\mathbf{h'}}]_j}\right\}=[\mathbf{J}^D_{\mathbf{y}^R_\mathcal{Q}}(\tilde{\mathbf{h'}})]_{ij}+[\mathbf{J}^D_{\mathbf{y}^I_\mathcal{Q}}(\tilde{\mathbf{h'}})]_{ij}.
    \label{equ_Jdd}
 \end{equation}
With the assumption $\tilde{\mathbf{h'}}$ is Gaussian distributed with zero mean and covariance matrix $\mathbf{C}_{\tilde{\mathbf{h'}}}=\frac{1}{2}\mathbf{I}_2\otimes\mathbf{C}_\mathbf{h'}$, $\ln p(\tilde{\mathbf{h'}})$ yields
\begin{equation}
    \ln p(\tilde{\mathbf{h'}}) = -\frac{1}{2}N_rN_t\ln\left[(2\pi)^{2N_rN_t}\det(\mathbf{C}_{\tilde{\mathbf{h'}}})\right]-\frac{1}{2}\tilde{\mathbf{h'}}^T\mathbf{C}_{\tilde{\mathbf{h'}}}^{-1}\tilde{\mathbf{h'}}
\end{equation}
and inserted into (\ref{equ_h}) we obtain
\begin{equation}
    \mathbf{J}^P_{\mathbf{y}_\mathcal{Q}}(\tilde{\mathbf{h'}})=2\mathbf{J}^P_{\mathbf{y}^{R/I}_\mathcal{Q}}(\tilde{\mathbf{h'}})=2\mathbf{C}_{\tilde{\mathbf{h'}}}^{-1}.
    \label{equ_Jp}
\end{equation}
The BIM is the summation of (\ref{equ_Jdd}) and (\ref{equ_Jp}) as described by
\begin{equation} \mathbf{J}_{\mathbf{y}_\mathcal{Q}}(\tilde{\mathbf{h'}})=\mathbf{J}^D_{\mathbf{y}_\mathcal{Q}}(\tilde{\mathbf{h'}})+\mathbf{J}^P_{\mathbf{y}_\mathcal{Q}}(\tilde{\mathbf{h'}}).
\end{equation}

\subsubsection{BIM for Oversampled Systems}
When $M\geq2$ the equivalent noise vector $\mathbf{n}$ consists of colored Gaussian noise samples. Computing $p(\mathbf{y}_\mathcal{Q}^{R/I}\mid\tilde{\mathbf{h'}})$ requires the orthant probabilities, which are not available or too difficult to compute. The authors in \cite{6783980} have given a lower bound of $\mathbf{J}^D_{\mathbf{y}_\mathcal{Q}^{R/I}}(\tilde{\mathbf{h'}})$, which is based on the first and second order moments
\begin{equation}
    \mathbf{J}^D_{\mathbf{y}_\mathcal{Q}^{R/I}}(\tilde{\mathbf{h'}})\geq\left(\frac{\partial \mathbf{\mu}_{\mathbf{y}_\mathcal{Q}^{R/I}}}{\partial \tilde{\mathbf{h'}}}\right)^T\mathbf{C}^{-1}_{\mathbf{y}^{R/I}_\mathcal{Q}}\left(\frac{\partial \mathbf{\mu}_{\mathbf{y}_\mathcal{Q}^{R/I}}}{\partial \tilde{\mathbf{h'}}}\right) = \tilde{\mathbf{J}}^D_{\mathbf{y}_\mathcal{Q}^{R/I}}(\tilde{\mathbf{h'}}).
\end{equation}
Since the lower-bounding technique is identical to the real and the imaginary part, we present only the derivation of $\tilde{\mathbf{J}}^D_{\mathbf{y}_\mathcal{Q}^R}(\tilde{\mathbf{h'}})$. Based on \cite{8445905} \cite{Stein2017}, the mean value of the $k$th received symbol is given by
\begin{equation}
\begin{aligned}
[\mathbf{\mu}_{\mathbf{y}_\mathcal{Q}^R}]_k &= \frac{1}{\sqrt{2}}p\left([\mathbf{y}_\mathcal{Q}^R]_k=+1\mid\tilde{\mathbf{h'}}\right)-\frac{1}{\sqrt{2}}p\left([\mathbf{y}_\mathcal{Q}^R]_k=-1\mid\tilde{\mathbf{h'}}\right)\\&=\frac{1}{\sqrt{2}}\left[1-2Q\left(\frac{[\mathbf{\Phi}^R\mathbf{h'}^R-\mathbf{\Phi}^I\mathbf{h'}^I]_k}{\sqrt{[\mathbf{C}_\mathbf{n}]_{kk}/2}}\right)\right],
\end{aligned}
\label{equ_miu}
\end{equation}
The derivative of (\ref{equ_miu}) is
\begin{equation}
\frac{\partial [\mathbf{\mu}_{\mathbf{y}_\mathcal{Q}^R}]_k}{\partial [\tilde{\mathbf{h'}}]_i}=\frac{2\text{exp}\left(-\frac{[\mathbf{\Phi}^R\mathbf{h'}^R-\mathbf{\Phi}^I\mathbf{h'}^I]_k^2}{[\mathbf{C}_\mathbf{n}]_{kk}}\right)\frac{\partial [\mathbf{\Phi}^R\mathbf{h'}^R-\mathbf{\Phi}^I\mathbf{h'}^I]_k}{\partial [\tilde{\mathbf{h'}}]_i}}{\sqrt{2\pi[\mathbf{C}_\mathbf{n}]_{kk}}}.
\end{equation}
The diagonal elements of the covariance matrix are given by
\begin{equation}
[\mathbf{C}_{\mathbf{y}^R_\mathcal{Q}}]_{kk}=\frac{1}{2}-[\mathbf{\mu}_{\mathbf{y}_\mathcal{Q}^R}]_k^2,
\end{equation}
while the off-diagonal elements are calculated as
\begin{equation}
\begin{aligned}
[\mathbf{C}_{\mathbf{y}^R_\mathcal{Q}}]_{kn}&=p(z_k>0,z_n>0)+p(z_k\leq0,z_n\leq0)\\&\hspace{2.5cm}-\frac{1}{2}-[\mathbf{\mu}_{\mathbf{y}_\mathcal{Q}^R}]_k[\mathbf{\mu}_{\mathbf{y}_\mathcal{Q}^R}]_n,
\end{aligned}
\end{equation}
where $[z_k,z_n]^T$ is a bi-variate Gaussian random vector
\begin{equation*}
\begin{bmatrix}
z_k\\z_n
\end{bmatrix}\sim\mathcal{N}\left(\begin{bmatrix}
[\mathbf{\Phi}^R\mathbf{h'}^R-\mathbf{\Phi}^I\mathbf{h'}^I]_k\\ [\mathbf{\Phi}^R\mathbf{h'}^R-\mathbf{\Phi}^I\mathbf{h'}^I]_n
\end{bmatrix},\frac{1}{2}\begin{bmatrix}
[\mathbf{C}_\mathbf{n}]_{kk} & [\mathbf{C}_\mathbf{n}]_{kn}\\
[\mathbf{C}_\mathbf{n}]_{nk} & [\mathbf{C}_\mathbf{n}]_{nn}
\end{bmatrix}\right).
\end{equation*}
The lower bound for the imaginary part is derived in the same way. With the calculations above we get the lower bound of the BIM as
\begin{equation}
    \mathbf{J}_{\mathbf{y}_\mathcal{Q}}(\tilde{\mathbf{h'}})\geq \tilde{\mathbf{J}}^D_{\mathbf{y}_\mathcal{Q}}(\tilde{\mathbf{h'}})+\mathbf{J}^P_{\mathbf{y}_\mathcal{Q}}(\tilde{\mathbf{h'}}),
\end{equation}
where the equality holds for $M=1$ \cite{6783980}\cite{8445905}. Based on (21), the inverse of this BIM lower bound will result in an upper bound of the actual Bayesian CRB for the oversampled systems.

\subsection{Oversampling based LRA-LMMSE Channel Estimation}
In the uplink transmission phase, each block can be divided into two parts: one for training and the other for data transmission. During the training, all terminals simultaneously transmit their pilot sequences of $\tau$ symbols to the BS, which yields
\begin{equation}
    \mathbf{y}_{\mathcal{Q}_p}= \mathcal{Q}(\mathbf{\Phi}_p\mathbf{h'} + \mathbf{n}_p)=\tilde{\mathbf{\Phi}}\mathbf{h'} + \tilde{\mathbf{n}}_p,
    \label{equ_linear}
\end{equation}
where $\tilde{\mathbf{\Phi}} = \mathbf{A}_p\mathbf{\Phi}_p$ and $\tilde{\mathbf{n}}_p = \mathbf{A}_p\mathbf{n}_p+\mathbf{n}_q$. The vector $\mathbf{n}_q$ is the statistically equivalent quantizer noise. The matrix $\mathbf{A}_p$ is the Bussgang based linear operator chosen independently from $\mathbf{y}_p$:
\begin{equation}
\mathbf{A}_p=\mathbf{C}_{\mathbf{y}_p\mathbf{y}_{\mathcal{Q}_p}}^H\mathbf{C}_{\mathbf{y}_p}^{-1}=\sqrt{\frac{2}{\pi}}\text{diag}\left(\mathbf{C}_{\mathbf{y}_p}\right)^{-\frac{1}{2}},
\end{equation}
where $\mathbf{C}_{\mathbf{y}_p\mathbf{y}_{\mathcal{Q}_p}}=E\{\mathbf{y}_p\mathbf{y}^H_{\mathcal{Q}_p}\}$ denotes the cross-correlation matrix between the received signal $\mathbf{y}_p$ and the quantized signal $\mathbf{y}_{\mathcal{Q}_p}$. $\mathbf{C}_{\mathbf{y}_p}=E\{\mathbf{y}_p\mathbf{y}_p^H\}$ is the auto-correlation matrix of $\mathbf{y}_p$ given by
\begin{equation}
\mathbf{C}_{\mathbf{y}_p}=\mathbf{\Phi}_p\mathbf{C}_\mathbf{h'}\mathbf{\Phi}_p^H+\sigma_n^2\mathbf{I}_{N_r}\otimes \mathbf{GG}^H.
\label{equ_Cy}
\end{equation}
Based on the statistically equivalent linear model (\ref{equ_linear}), the proposed oversampling based LRA-LMMSE channel estimator is given by
\begin{equation}
\hat{\mathbf{h'}}_{\text{LMMSE}} = \mathbf{C}_\mathbf{h'}\tilde{\mathbf{\Phi}}^H\mathbf{C}_{\mathbf{y}_{\mathcal{Q}_p}}^{-1}\mathbf{y}_{\mathcal{Q}_p}.
\label{equ_blmmse}
\end{equation}
Note that when $M=1$, (\ref{equ_blmmse}) reduces to the same as that
of the BLMMSE channel estimator in \cite{7931630}. Other channel
estimation techniques that exploit low-rank and other recursive
strategies can also be pursued \cite{jio,jidf,jiotvt}.

\section{Numerical Results}

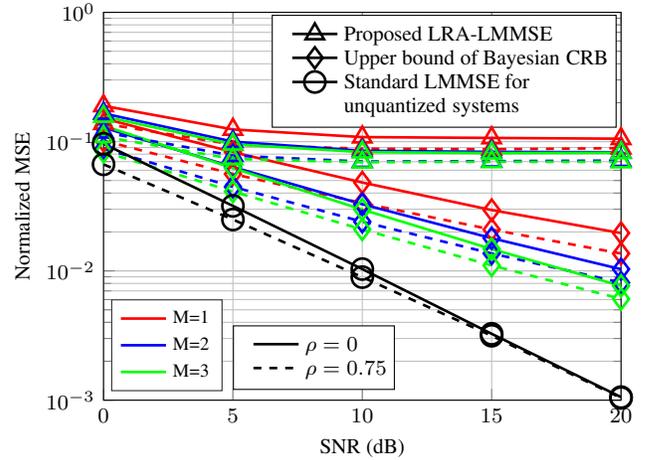
\begin{figure}[!htbp]
    \centering
    \input{MSE.tex}
    \caption{$N_t = 4$ and $N_r = 16$. Normalized MSE comparison as a function of $\text{SNR}$ when $\tau = 40$.}
    \label{fig:MSE}
\end{figure}

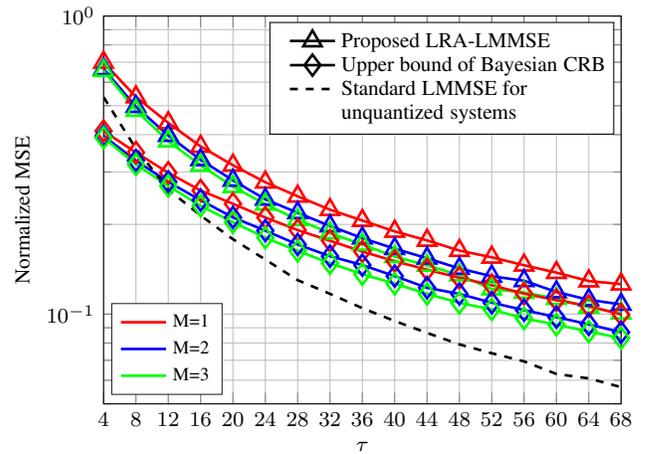
\begin{figure}[!htbp]
    \centering
    \input{MSE_pilots.tex}
    \caption{$N_t = 4$ and $N_r = 16$. Normalized MSE comparison as a function of $\tau$ when $\text{SNR}=0\text{dB}$ and $\rho=0$.}
    \label{fig:MSE_pilots}
\end{figure}

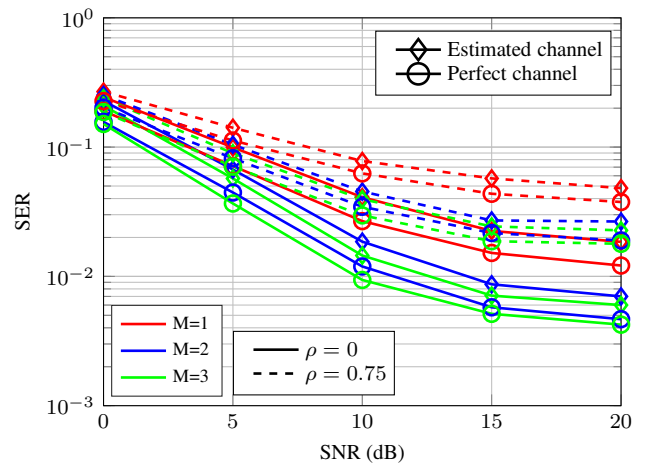
\begin{figure}[!htbp]
    \centering
    \input{SER.tex}
    \caption{$N_t = 4$ and $N_r = 16$. SER comparison for different oversampling factors when $\tau = 40$.}
    \label{fig:SER}
\end{figure}

This section presents simulation results of the proposed LRA-LMMSE
channel estimaton. The modulation scheme is QPSK. The $m(t)$ and
$p(t)$ are normalized Root-Raised-Cosine (RRC) filters with a
roll-off factor of 0.8. The channel is assumed to experience block
fading and is modeled as the Kronecker model \cite{837052}
$\mathbf{H'} =
\mathbf{R}_r^{\frac{1}{2}}\mathbf{H}'_w\mathbf{R}_t^{\frac{1}{2}}$,
where $\mathbf{R}_r$ and $\mathbf{R}_t$ denote the receive and
transmit correlation matrices, respectively. The elements of
$\mathbf{H}'_w$ are i.i.d. complex Gaussian random variables with
zero mean and unit variance.  $\mathbf{R}_t = \mathbf{I}_{N_t}$ by
assuming that the channel of each terminal is independent. The
$\mathbf{R}_r$ has the following form:
\begin{equation}
\mathbf{R}_r = \begin{bmatrix}
1 & \rho & \dots & \rho^{(N_r-1)^2}\\
\rho & 1 & \dots & \rho^{(N_r-2)^2}\\
\vdots & \vdots & \ddots & \vdots\\
\rho^{(N_r-1)^2} & \rho^{(N_r-2)^2} & \dots & 1\\
\end{bmatrix},
\end{equation}
where $\rho$ is the correlation index of neighboring antennas
($\rho$ = 0 represents an uncorrelated scenario and $\rho$ = 1
implies a fully correlated scenario). The pilots are column-wise
orthogonal. The signal-to-noise ratio (SNR) is defined as
$10\log(\frac{N_t}{\sigma_n^2})$. The normalized mean square error
(MSE) performances are illustrated in Fig.\ref{fig:MSE}, where there
is a 2 dB performance gain of the proposed oversampling based
LRA-LMMSE channel estimator compared to the BLMMSE channel estimator
($M=1$). Note that for the oversampled systems ($M\geq2$) the upper
bound of Bayesian CRBs are higher than the actual Bayesian CRBs,
since the calculation of the actual Bayesian CRBs are still open
problems. Fig.\ref{fig:MSE_pilots} shows the normalized MSE
performances as a function of pilot symbols $\tau$. To achieve a
trade-off between MSE performance and system complexity we have set
$\tau=40$ in the simulation. Moreover, the symbol error rate (SER)
performances of the system with the proposed LRA-LMMSE channel
estimator and perfect channel matrix are shown in Fig.\ref{fig:SER},
where the sliding-window based LMMSE detector \cite{8450809} with
window length three is applied in the system. Further investigation
with nonlinear detectors
\cite{itic,spa,mfsic,mfdf,mbdf,did,bfidd,8240730} will be considered
elsewhere.

\section{Conclusion}
This work has proposed the oversampling based LRA-LMMSE channel
estimator for uplink massive MIMO systems with 1-bit quantization
and oversampling at the receiver. We have further given analytical
performance of the system in terms of the Bayesian information.
Simulation results have shown that the proposed oversampling based
channel estimator outperforms the existing non-oversampled BLMMSE
channel estimator in terms of the MSE and the SER performances.

\bibliographystyle{IEEEbib}
\bibliography{ref}

\end{document}

%% file: system_model.tex
\def\antenna{
	-- +(0mm,2.0mm) -- +(1.625mm,4.5mm) -- +(-1.625mm,4.5mm) -- +(0mm,2.0mm)
}
\tikzset{%
	harddecision/.style={draw, 
		path picture={
			\pgfpointdiff{\pgfpointanchor{path picture bounding box}{north east}}%
			{\pgfpointanchor{path picture bounding box}{south west}}
			\pgfgetlastxy\x\y
			\tikzset{x=\x*.4, y=\y*.4}
			%
			\draw (-0.5,-0.5)--(0,-0.5)--(0,0.5)--(0.5,0.5);  
			\draw (-0.25,0)--(0.25,0);
	}}
}

\begin{tikzpicture}
	\node (c0) {\footnotesize Terminal 1};
	\node[dspsquare, right= 0.5cm of c0,minimum width=1.4cm,text height=0.8em]       (c2) {\footnotesize Modulator};
	\node[dspsquare, right= 0.6cm of c2,minimum width=1.4cm]       (c10) {\footnotesize $p(t)$};
	\node[coordinate,right= 0.5cm of c10] (c3) {};

	\node[below= 0.25cm of c0] (c222) {\tiny \textbullet};
	\node[below= 0.04cm of c222] (c2222) {\tiny \textbullet};
	\node[below= 0.04cm of c2222] (c22222) {\tiny \textbullet};
	\node[coordinate,below= 0.62cm of c2] (cfix) {};
	
	\node[below= 1.6cm of c0] (c00) {\footnotesize Terminal $N_t$};
	\node[dspsquare, right= 0.4cm of c00,minimum width=1.4cm,text height=0.8em]                    (c22) {\footnotesize Modulator};
	\node[dspsquare, right= 0.6cm of c22,minimum width=1.4cm]       (c12) {\footnotesize $p(t)$};
	\node[coordinate,right= 0.5cm of c12] (c33) {};
	
	\node[coordinate,right= 0.7cm of c3] (c8) {};
	\node[coordinate,right= 0.7cm of c33] (c9) {};
	
	\node[coordinate,right= 0.5cm of c10.-15] (c88) {};
	\node[coordinate,below= 1.2cm of c88] (c99) {};
	
	\node[dspsquare, right= 0.5cm of c8,minimum width=1.4cm]       (c42) {\footnotesize $m(t)$};
	\node[dspsquare, right= 0.5cm of c9,minimum width=1.4cm]       (c43) {\footnotesize $m(t)$};
	
	\node[dspsquare, right= 0.5cm of c42,minimum width=2cm,text height=2em]       (c13) {\footnotesize M-fold \\ Oversampling};
	\node[dspsquare, right= 0.5cm of c43,minimum width=2cm,text height=2em]       (c14) {\footnotesize M-fold \\ Oversampling};
	
	\node[dspsquare,right= 0.7cm of c13,minimum width=0.8cm,text height=0.8em] (c15) {\footnotesize $\mathcal{Q}(\cdot)$};
	\node[dspsquare,right= 0.7cm of c14,minimum width=0.8cm,text height=0.8em] (c16) {\footnotesize $\mathcal{Q}(\cdot)$};
	
%
	
	\node[dspsquare,right= 10.7cm of cfix,minimum height=3cm,text height=2.5em,minimum width=1.2cm] (c17) {\footnotesize Channel \\\footnotesize Estimator };
	
	\node[coordinate,right= 1cm of c17] (c20) {};
	
	\node[coordinate,right= 0.65cm of c88] (c200) {};
	\node[coordinate,right= 0.65cm of c99] (c211) {};
	
	\draw[dspconn] (c0) -- node[] {} (c2);
	\draw[dspconn] (c2) -- node[midway,below] {\footnotesize $x_1$} (c10);
	\draw[dspconn] (c2) -- node[midway,above] {\footnotesize $\frac{1}{T}$} (c10);
	\draw[thick] (c10) -- node[] {} (c3);
	\draw[dspconn] (c00) -- node[] {} (c22);
	\draw[dspconn] (c22) -- node[midway,below] {\footnotesize $x_{N_t}$} (c12);
	\draw[dspconn] (c22) -- node[midway,above] {\footnotesize $\frac{1}{T}$} (c12);
	\draw[thick] (c12) -- node[] {} (c33);
	\draw [thick] (c3) \antenna;
	\draw [thick] (c33) \antenna;
	\draw [thick] (c8) \antenna;
	\draw [thick] (c9) \antenna;
	\draw[densely dotted] (c88) -- node[] {} (c211);
	\draw[densely dotted] (c99) -- node[] {} (c200);
	\draw[thick] (c8) -- node[] {} (c42);
	\draw[thick] (c9) -- node[] {} (c43);
	\draw[thick] (c42) -- node[midway,above] {} (c13);
	\draw[thick] (c43) -- node[midway,above] {} (c14);
	\draw[dspconn] (c13) -- node[midway,above] {\footnotesize $\frac{M}{T}$} (c15);
	\draw[dspconn] (c14) -- node[midway,above] {\footnotesize $\frac{M}{T}$} (c16);
	\draw[dspconn] (c13) -- node[midway,below] {\footnotesize $y_1$} (c15);
	\draw[dspconn] (c14) -- node[midway,below] {\footnotesize $y_{N_r}$} (c16);
	\draw[dspconn] (c15) -- node[midway,below] {\footnotesize $y_{\mathcal{Q}_1}$} (c17.121);
	\draw[dspconn] (c16) -- node[midway,below] {\footnotesize $y_{\mathcal{Q}_{N_r}}$} (c17.-121);
	\draw[dspconn] (c17) -- node[midway,above] {\footnotesize $\mathbf{\hat{h}}$} (c20);

	\end{tikzpicture}

%% file: MSE.tex
\pgfplotsset{every axis label/.append style={font=\footnotesize},
	every tick label/.append style={font=\footnotesize},
}

\begin{tikzpicture}

\begin{axis}[%
width=.8\columnwidth,
height=.6\columnwidth,
at={(0.758in,0.603in)},
scale only axis,
xmin=0,
xmax=20,
xlabel style={font=\footnotesize},
xlabel={SNR (dB)},
xtick=data,
xmajorgrids,
ymin=0.001,
ymax=1,
yminorticks=true,
ymode=log,
ylabel style={font=\footnotesize},
ylabel={Normalized MSE},
ymajorgrids,
yminorgrids,
axis background/.style={fill=white},
legend entries={M=1,
	M=2,
	M=3},
legend style={at={(0.23,0.26)},anchor=north east,legend cell align=left,align=left,draw=white!15!black,font=\scriptsize}
]

\addlegendimage{color=red,fill=gray!20,line width=1.0pt,mark size=4pt}
\addlegendimage{color=blue,fill=green!20,line width=1.0pt,mark size=4pt}
\addlegendimage{color=green,fill=gray!20,line width=1.0pt,mark size=4pt}

\addplot [color=red,solid,line width=1.0pt,mark=diamond,mark options={solid},mark size=4pt]
table[row sep=crcr]{%
	0	0.151128375028731\\
	5	0.0829411608298734\\
	10	0.0482509743429984\\
	15	0.0295511924760972\\
	20  0.0196132854654152\\
};

\addplot [color=red,dashed,line width=1.0pt,mark=diamond,mark options={solid},mark size=4pt]
table[row sep=crcr]{%
	0	0.101688004580020\\
	5	0.0561925314876146\\
	10	0.0334681307629001\\
	15	0.0208750508948124\\
	20  0.0136261609860353\\
};

\addplot [color=blue,solid,line width=1.0pt,mark=diamond,mark options={solid},mark size=4pt]
  table[row sep=crcr]{%
  	0	0.128535728738479\\
  	5	0.0636058046760995\\
  	10	0.0328023452189701\\
  	15  0.0179054969237334\\
  	20	0.0103065594044048\\
};

\addplot [color=blue,dashed,line width=1.0pt,mark=diamond,mark options={solid},mark size=4pt]
table[row sep=crcr]{%
	0	0.0881846941973652\\
	5	0.0446784130393671\\
	10	0.0240214658019199\\
	15  0.0136430331295080\\
	20	0.00811802030537684\\
};

\addplot [color=green,solid,line width=1.0pt,mark=diamond,mark options={solid},mark size=4pt]
table[row sep=crcr]{%
	0	0.132753536194842\\
	5	0.0623883497383276\\
	10  0.0298274944573941\\
	15	0.0147613167790440\\
	20	0.00765428932051721\\
};

\addplot [color=green,dashed,line width=1.0pt,mark=diamond,mark options={solid},mark size=4pt]
table[row sep=crcr]{%
	0	0.0836296922516933\\
	5	0.0410657070627449\\
	10	0.0209236425888670\\
	15	0.0110225375249120\\
	20	0.00607384735913191\\
};

\addplot [color=red,solid,line width=1.0pt,mark=triangle,mark options={solid},mark size=4pt]
table[row sep=crcr]{%
	0	0.189225348781723\\
	5	0.124181323568436\\
	10	0.108335400182167\\
	15	0.106748775446626\\
	20  0.105270902592677\\
};

\addplot [color=red,dashed,line width=1.0pt,mark=triangle,mark options={solid},mark size=4pt]
table[row sep=crcr]{%
	0	0.137697513333319\\
	5	0.0949036811703384\\
	10	0.0881426469685018\\
	15	0.0871815617898476\\
	20  0.0890252951376361\\
};

\addplot [color=blue,solid,line width=1.0pt,mark=triangle,mark options={solid},mark size=4pt]
table[row sep=crcr]{%
	0   0.165264026714237\\
	5	0.0999267285779435\\
	10	0.0849092437514610\\
	15  0.0833140420645897\\
	20  0.0827521320265596\\
};

\addplot [color=blue,dashed,line width=1.0pt,mark=triangle,mark options={solid},mark size=4pt]
table[row sep=crcr]{%
	0   0.118116649182904\\
	5	0.0779820794776814\\
	10	0.0700743126663910\\
	15  0.0706259916194245\\
	20  0.0712749117688770\\
};

\addplot [color=green,solid,line width=1.0pt,mark=triangle,mark options={solid},mark size=4pt]
table[row sep=crcr]{%
	0	0.1566\\
	5	0.0954\\
	10	0.0829\\
	15	0.0815\\
	20	0.0820\\
};

\addplot [color=green,dashed,line width=1.0pt,mark=triangle,mark options={solid},mark size=4pt]
table[row sep=crcr]{%
	0	0.109170041503051\\
	5	0.0735011403583949\\
	10	0.0692128942364387\\
	15	0.0698031124199272\\
	20	0.0696123916841031\\
};

\addplot [color=black,solid,line width=1.0pt,mark=o,mark options={solid},mark size=4pt]
table[row sep=crcr]{%
	0	0.0961861872586514\\
	5	0.0318247246853578\\
	10	0.0103282524339940\\
	15	0.00326071290290002\\
	20	0.00105618368971160\\
};

\addplot [color=black,dashed,line width=1.0pt,mark=o,mark options={solid},mark size=4pt]
table[row sep=crcr]{%
	0	0.0667072853003805\\
	5	0.0249593705784238\\
	10	0.00894927336059394\\
	15	0.00314876621218350\\
	20	0.00104010991851947\\
};

\addplot[only marks,smooth,color=black,solid,line width=1.0pt,mark=diamond,mark size=4pt,
y filter/.code={\pgfmathparse{\pgfmathresult-0}\pgfmathresult}]
table[row sep=crcr]{%
	1 2 3 4 5\\
};\label{L21}

\addplot[only marks,smooth,color=black,solid,line width=1.0pt,mark=triangle,mark size=4pt,
y filter/.code={\pgfmathparse{\pgfmathresult-0}\pgfmathresult}]
table[row sep=crcr]{%
	1 2 3 4 5\\
};\label{L22}

\addplot[only marks,smooth,color=black,solid,line width=1.0pt,mark=o,mark size=4pt,y filter/.code={\pgfmathparse{\pgfmathresult-0}\pgfmathresult}]
table[row sep=crcr]{%
	1 2 3 4 5\\
};\label{L23}

\addplot[smooth,color=black,solid,line width=1.0pt,
y filter/.code={\pgfmathparse{\pgfmathresult-0}\pgfmathresult}]
table[row sep=crcr]{%
	1 2 3 4 5\\
};\label{L22221}

\addplot[smooth,color=black,dashed,line width=1.0pt,
y filter/.code={\pgfmathparse{\pgfmathresult-0}\pgfmathresult}]
table[row sep=crcr]{%
	1 2 3 4 5\\
};\label{L22222}

\node [draw,fill=white,font=\footnotesize,anchor= south west,at={(6.5,0.128)}] {
	\setlength{\tabcolsep}{0mm}
	\renewcommand{\arraystretch}{1}
	\begin{tabular}{l}
	\ref{L22}{~Proposed LRA-LMMSE}\\
	\ref{L21}{~Upper bound of Bayesian CRB}\\
	\ref{L23}{~Standard LMMSE for}\\
	{~~~~~~~~~~unquantized systems}\\
	\end{tabular}
};

\node [draw,fill=white,font=\footnotesize,anchor= south west,at={(5,0.0012)}] {
	\setlength{\tabcolsep}{0mm}
	\renewcommand{\arraystretch}{1}
	\begin{tabular}{l}
	\ref{L22221}{~$\rho=0$}\\
	\ref{L22222}{~$\rho=0.75$}\\
	\end{tabular}
};

\end{axis}

\end{tikzpicture}%

%% file: MSE_pilots.tex
\pgfplotsset{every axis label/.append style={font=\footnotesize},
	every tick label/.append style={font=\footnotesize},
}

\begin{tikzpicture}

\begin{axis}[%
width=.8\columnwidth,
height=.6\columnwidth,
at={(0.758in,0.603in)},
scale only axis,
xmin=4,
xmax=68,
xlabel style={font=\footnotesize},
xlabel={$\tau$},
xtick=data,
xmajorgrids,
ymin=0.05,
ymax=1,
yminorticks=true,
ymode=log,
ylabel style={font=\footnotesize},
ylabel={Normalized MSE},
ymajorgrids,
yminorgrids,
axis background/.style={fill=white},
legend entries={M=1,
	M=2,
	M=3},
legend style={at={(0.23,0.26)},anchor=north east,legend cell align=left,align=left,draw=white!15!black,font=\scriptsize}
]

\addlegendimage{color=red,fill=gray!20,line width=1.0pt,mark size=4pt}
\addlegendimage{color=blue,fill=green!20,line width=1.0pt,mark size=4pt}
\addlegendimage{color=green,fill=gray!20,line width=1.0pt,mark size=4pt}

 M=1
\addplot [color=red,solid,line width=1.0pt,mark=triangle,mark options={solid},mark size=4pt]
table[row sep=crcr]{%
	4	0.699358873117348\\
	8	0.535387686085043\\
	12	0.438169912006865\\
	16	0.365300128827479\\
	20	0.315859033103829\\
	24	0.277005930784225\\
	28	0.248471095323321\\
	32	0.224113927055519\\
	36	0.206053229340759\\
	40	0.189507030164755\\
	44  0.176734347993079\\
	48	0.163415739671806\\
	52	0.155116975504648\\
	56  0.1459\\
	60  0.1378\\
	64  0.1291\\
	68  0.1261\\
};


\addplot [color=blue,solid,line width=1.0pt,mark=triangle,mark options={solid},mark size=4pt]
table[row sep=crcr]{%
	4	0.664749505617734\\
	8	0.496771912733656\\
	12	0.397437623800126\\
	16	0.329404638534690\\
	20	0.281070828100532\\
	24	0.242813606088337\\
	28	0.218800855336421\\
	32	0.197575139778541\\
	36	0.179906383400995\\
	40	0.165807822317055\\
	44	0.154012424378742\\
	48  0.142315629379899\\
	52	0.134009996726074\\
	56	0.1294\\
	60	0.1187\\
	64  0.1117\\
	68  0.1078\\
};


\addplot [color=green,solid,line width=1.0pt,mark=triangle,mark options={solid},mark size=4pt]
table[row sep=crcr]{%
	4	0.655976762812469\\
	8	0.482514795711538\\
	12	0.380500014623567\\
	16	0.315806863098145\\
	20	0.267725981503807\\
	24	0.233595165040051\\
	28	0.207355299012001\\
	32	0.187077472863596\\
	36	0.171270479785475\\
	40	0.155938079611792\\
	44	0.145683323380260\\
	48	0.135266278096005\\
	52	0.1214\\
	56	0.118206172977920\\
	60	0.1127\\
	64	0.1055\\
	68  0.1008\\
};

\addplot [color=red,solid,line width=1.0pt,mark=diamond,mark options={solid},mark size=4pt]
table[row sep=crcr]{%
	4	0.411405696477307\\
	8	0.348637394838640\\
	12	0.297960547069896\\
	16	0.259728182249572\\
	20	0.234519779112624\\
	24	0.210910032630642\\
	28	0.191443740257952\\
	32	0.176257119658564\\
	36	0.162166751918272\\
	40	0.150549374372306\\
	44  0.141026843819590\\
	48	0.132429079257141\\
	52	0.125046404600826\\
	56	0.117147041951399\\
	60	0.111410590644421\\ 
	64	0.106443540580195\\
	68	0.0996910667660181\\
};


\addplot [color=blue,solid,line width=1.0pt,mark=diamond,mark options={solid},mark size=4pt]
table[row sep=crcr]{%
	4	0.396130863710951\\
	8	0.327469940422354\\
	12	0.277702578739516\\
	16	0.241057425736248\\
	20	0.211095142973932\\
	24	0.190627730753693\\
	28	0.171228299133600\\
	32	0.156449751139604\\
	36	0.146082876280077\\
	40	0.134178093252914\\
	44  0.122598683193984\\
	48	0.116874817796306\\
	52	0.109272872126709\\
	56	0.103154731143297\\
	60  0.0977404987890549\\
	64  0.0924269065620093\\
	68	0.0869010924170605\\
};

\addplot [color=green,solid,line width=1.0pt,mark=diamond,mark options={solid},mark size=4pt]
table[row sep=crcr]{%
	4	0.3909\\
	8	0.3180\\
	12	0.2693\\
	16	0.2309\\
	20	0.2034\\
	24	0.1805\\
	28	0.1631\\
	32	0.1482\\
	36	0.1362\\
	40	0.1265\\
	44  0.1170\\
	48	0.1091\\
	52	0.1037\\
	56  0.0967\\
	60  0.0922\\
	64  0.0874\\
	68  0.0831\\
};

\addplot [color=black,dashed,line width=1.0pt]
table[row sep=crcr]{%
	4	0.534581777625582\\
	8	0.359190381211243\\
	12	0.261629718940548\\
	16	0.214206569786776\\
	20	0.178365598325460\\
	24	0.152466970469423\\
	28	0.130035479543129\\
	32	0.116982752025296\\
	36	0.104736530561434\\
	40	0.0948394183783961\\
	44  0.0864942132590994\\
	48	0.0791634856760853\\
	52	0.0739442892445326\\
	56	0.0694235633211098\\
	60	0.0630001756065135\\
	64	0.0608635894053683\\
	68	0.0568909623299040\\
};


\addplot[only marks,smooth,color=black,solid,line width=1.0pt,mark=diamond,mark size=4pt,
y filter/.code={\pgfmathparse{\pgfmathresult-0}\pgfmathresult}]
table[row sep=crcr]{%
	1 2\\
};\label{L221}

\addplot[only marks,smooth,color=black,solid,line width=1.0pt,mark=triangle,mark size=4pt,
y filter/.code={\pgfmathparse{\pgfmathresult-0}\pgfmathresult}]
table[row sep=crcr]{%
	1 2\\
};\label{L222}

\addplot[smooth,color=black,dashed,line width=1.0pt,
y filter/.code={\pgfmathparse{\pgfmathresult-0}\pgfmathresult}]
table[row sep=crcr]{%
	1 2 3\\
};\label{L223}

\node [draw,fill=white,font=\footnotesize,anchor= south west,at={(24.5,0.4)}] {
	\setlength{\tabcolsep}{0mm}
	\renewcommand{\arraystretch}{1}
	\begin{tabular}{l}
	\ref{L222}{~Proposed LRA-LMMSE}\\
	\ref{L221}{~Upper bound of Bayesian CRB}\\
	\ref{L223}{~Standard LMMSE for}\\
	{~~~~~~~~~~unquantized systems}\\
	\end{tabular}
};

\end{axis}

\end{tikzpicture}%

%% file: SER.tex
\pgfplotsset{every axis label/.append style={font=\footnotesize},
	every tick label/.append style={font=\footnotesize},
}

\begin{tikzpicture}

\begin{axis}[%
width=.8\columnwidth,
height=.6\columnwidth,
at={(0.758in,0.603in)},
scale only axis,
xmin=0,
xmax=20,
xlabel style={font=\footnotesize},
xlabel={SNR (dB)},
xtick=data,
xmajorgrids,
ymode=log,
ymin=1e-3,
ymax=1,
yminorticks=true,
ylabel style={font=\footnotesize},
ylabel={SER},
ymajorgrids,
yminorgrids,
axis background/.style={fill=white},
legend entries={M=1,
	M=2,
	M=3},
legend style={at={(0.23,0.26)},anchor=north east,legend cell align=left,align=left,draw=white!15!black,font=\scriptsize}
]

\addlegendimage{color=red,fill=gray!20,line width=1.0pt,mark size=4pt}
\addlegendimage{color=blue,fill=green!20,line width=1.0pt,mark size=4pt}
\addlegendimage{color=green,fill=gray!20,line width=1.0pt,mark size=4pt}

\addplot [color=red,solid,line width=1.0pt,mark=diamond,mark options={solid},mark size=3pt]
table[row sep=crcr]{%
	0	0.243383333333333\\
	5	0.0994833333333333\\ 
	10	0.0409499999999999\\
	15	0.0224666666666666\\
	20	0.0185000000000000\\
};

\addplot [color=red,dashed,line width=1.0pt,mark=diamond,mark options={solid},mark size=3pt]
table[row sep=crcr]{%
	0	0.268483333333333\\
	5	0.141116666666667\\ 
	10	0.0780000000000000\\
	15	0.0572666666666667\\
	20	0.0481333333333333\\
};

\addplot [color=blue,solid,line width=1.0pt,mark=diamond,mark options={solid},mark size=3pt]
table[row sep=crcr]{%
	0	0.227600000000000\\
	5	0.0670833333333334\\ 
	10	0.0185666666666666\\
	15	0.00868333333333330\\
	20	0.00699999999999998\\
};

\addplot [color=blue,dashed,line width=1.0pt,mark=diamond,mark options={solid},mark size=3pt]
table[row sep=crcr]{%
	0	0.252583333333333\\
	5	0.104433333333333\\ 
	10	0.0453333333333333\\
	15	0.0270833333333333\\
	20	0.0265666666666666\\
};

\addplot [color=green,solid,line width=1.0pt,mark=diamond,mark options={solid},mark size=3pt]
table[row sep=crcr]{%
	0	0.2091\\
	5	0.0572\\ 
	10	0.0145\\
	15	0.0071\\
	20	0.0060\\
};

\addplot [color=green,dashed,line width=1.0pt,mark=diamond,mark options={solid},mark size=3pt]
table[row sep=crcr]{%
	0	0.238900000000000\\
	5	0.0895999999999999\\ 
	10	0.0395166666666666\\
	15	0.0243500000000000\\
	20	0.0227000000000000\\
};

\addplot [color=red,solid,line width=1.0pt,mark=o,mark options={solid},mark size=3pt]
table[row sep=crcr]{%
	0   0.190283333333333\\
	5	0.0714500000000000\\ 
	10	0.0267500000000000\\
	15	0.0151833333333333\\
	20	0.0121333333333333\\
};

\addplot [color=red,dashed,line width=1.0pt,mark=o,mark options={solid},mark size=3pt]
table[row sep=crcr]{%
	0   0.227050000000000\\
	5   0.112400000000000\\ 
	10	0.0626166666666666\\
	15	0.0435666666666666\\
	20	0.0376166666666667\\
};

\addplot [color=blue,solid,line width=1.0pt,mark=o,mark options={solid},mark size=3pt]
table[row sep=crcr]{%
	0	0.156766666666667\\
	5	0.0445500000000000\\ 
	10	0.0119333333333333\\
	15	0.00576666666666665\\
	20	0.00466666666666666\\
};

\addplot [color=blue,dashed,line width=1.0pt,mark=o,mark options={solid},mark size=3pt]
table[row sep=crcr]{%
	0	0.202050000000000\\
	5	0.0819500000000000\\ 
	10	0.0345666666666667\\
	15	0.0216499999999999\\
	20	0.0189666666666666\\
};

\addplot [color=green,solid,line width=1.0pt,mark=o,mark options={solid},mark size=3pt]
table[row sep=crcr]{%
	0	0.150950000000000\\
	5	0.0366333333333333\\ 
	10	0.00938333333333330\\
	15	0.00513333333333332\\
	20	0.00423333333333333\\
};

\addplot [color=green,dashed,line width=1.0pt,mark=o,mark options={solid},mark size=3pt]
table[row sep=crcr]{%
	0	0.186766666666667\\
	5	0.0707333333333333\\ 
	10	0.0296333333333333\\
	15	0.0187166666666666\\
	20	0.0178666666666666\\
};

\addplot[only marks,smooth,color=black,solid,line width=1.0pt,mark=diamond,mark size=4pt,
y filter/.code={\pgfmathparse{\pgfmathresult-0}\pgfmathresult}]
table[row sep=crcr]{%
	1 2\\
};\label{L2221}

\addplot[only marks,smooth,color=black,solid,line width=1.0pt,mark=o,mark size=4pt,
y filter/.code={\pgfmathparse{\pgfmathresult-0}\pgfmathresult}]
table[row sep=crcr]{%
	1 2\\
};\label{L2222}

\addplot[smooth,color=black,solid,line width=1.0pt,
y filter/.code={\pgfmathparse{\pgfmathresult-0}\pgfmathresult}]
table[row sep=crcr]{%
	1 2 3 4 5\\
};\label{L222221}

\addplot[smooth,color=black,dashed,line width=1.0pt,
y filter/.code={\pgfmathparse{\pgfmathresult-0}\pgfmathresult}]
table[row sep=crcr]{%
	1 2 3 4 5\\
};\label{L222222}

\node [draw,fill=white,font=\footnotesize,anchor= south west,at={(10.5,0.25)}] {
	\setlength{\tabcolsep}{0mm}
	\renewcommand{\arraystretch}{1}
	\begin{tabular}{l}
	\ref{L2221}{~Estimated channel}\\
	\ref{L2222}{~Perfect channel}\\
	\end{tabular}
};

\node [draw,fill=white,font=\footnotesize,anchor= south west,at={(5,0.0012)}] {
	\setlength{\tabcolsep}{0mm}
	\renewcommand{\arraystretch}{1}
	\begin{tabular}{l}
	\ref{L222221}{~$\rho=0$}\\
	\ref{L222222}{~$\rho=0.75$}\\
	\end{tabular}
};

\end{axis}
\end{tikzpicture}%